\documentclass[12pt,JPD]{iopart}
\usepackage{graphicx}
\usepackage{gensymb}
\usepackage{hyperref}

\begin{document}

\title[Oxide Molecular Beam Epitaxy for Quantum Materials]{Advances in Complex Oxide Quantum Materials Through New Approaches to Molecular Beam Epitaxy}

\author{Gaurab Rimal$^{1,2}$ and Ryan B. Comes$^1$}

\address{$^1$Auburn University, Department of Physics, Auburn, AL, USA}
\address{$^2$Western Michigan University, Department of Physics, Kalamazoo, MI, USA}
\ead{gaurab.rimal@wmich.edu}
\ead{ryan.comes@auburn.edu}
\vspace{10pt}
\begin{indented}
\item[\today]
\end{indented}

\begin{abstract}

Molecular beam epitaxy (MBE), a workhorse of the semiconductor industry, has progressed rapidly in the last few decades in the development of novel materials. Recent developments in condensed matter and materials physics have seen the rise of many novel quantum materials that require ultra-clean and high-quality samples for fundamental studies and applications. Novel oxide-based quantum materials synthesized using MBE have advanced the development of the field and materials. In this review, we discuss  the recent progress in new MBE techniques that have enabled synthesis of complex oxides that exhibit ``quantum" phenomena, including superconductivity and topological electronic states. We show how these techniques have produced breakthroughs in the synthesis of 4d and 5d oxide films and heterostructures that are of particular interest as quantum materials. These new techniques in MBE offer a bright future for the synthesis of ultra-high quality oxide quantum materials.

\end{abstract}

%
%
%
%
%

\section{Introduction}
Transition metal complex oxides have been a source of rich new physics for decades, dating back to the discovery of the high T$_{c}$ cuprate superconductors in the 1980s \cite{bednorz_possible_1986}, colossal magnetoresistance in manganites in the 1990s \cite{jin_thousandfold_1994}, multiferroicity in BiFeO$_{3}$ in the 2000s \cite{wang_epitaxial_2003}, and the topological Hall effect in iridates and ruthenates in the 2010s \cite{matsuno_interface-driven_2016}. These are just a smattering of the  many ``quantum" phenomena observed in complex oxides. However, as with any complex materials system, the field is rife with difficulties in repeatable synthesis of these materials. Careful synthesis and characterization is critical to decouple intrinsic physics in oxide materials from extrinsic effects that emerge due to growth-induced defects in these systems. Over the past 20 years, molecular beam epitaxy (MBE) of complex oxides has grown by leaps and bounds to become the state-of-the-art growth technique for epitaxial quantum oxide films. For details on the history of oxide MBE, we refer the readers to past review articles \cite{brahlek_frontiers_2018, schlom_perspective_2015}. 

\begin{figure}
    \centering
    \includegraphics{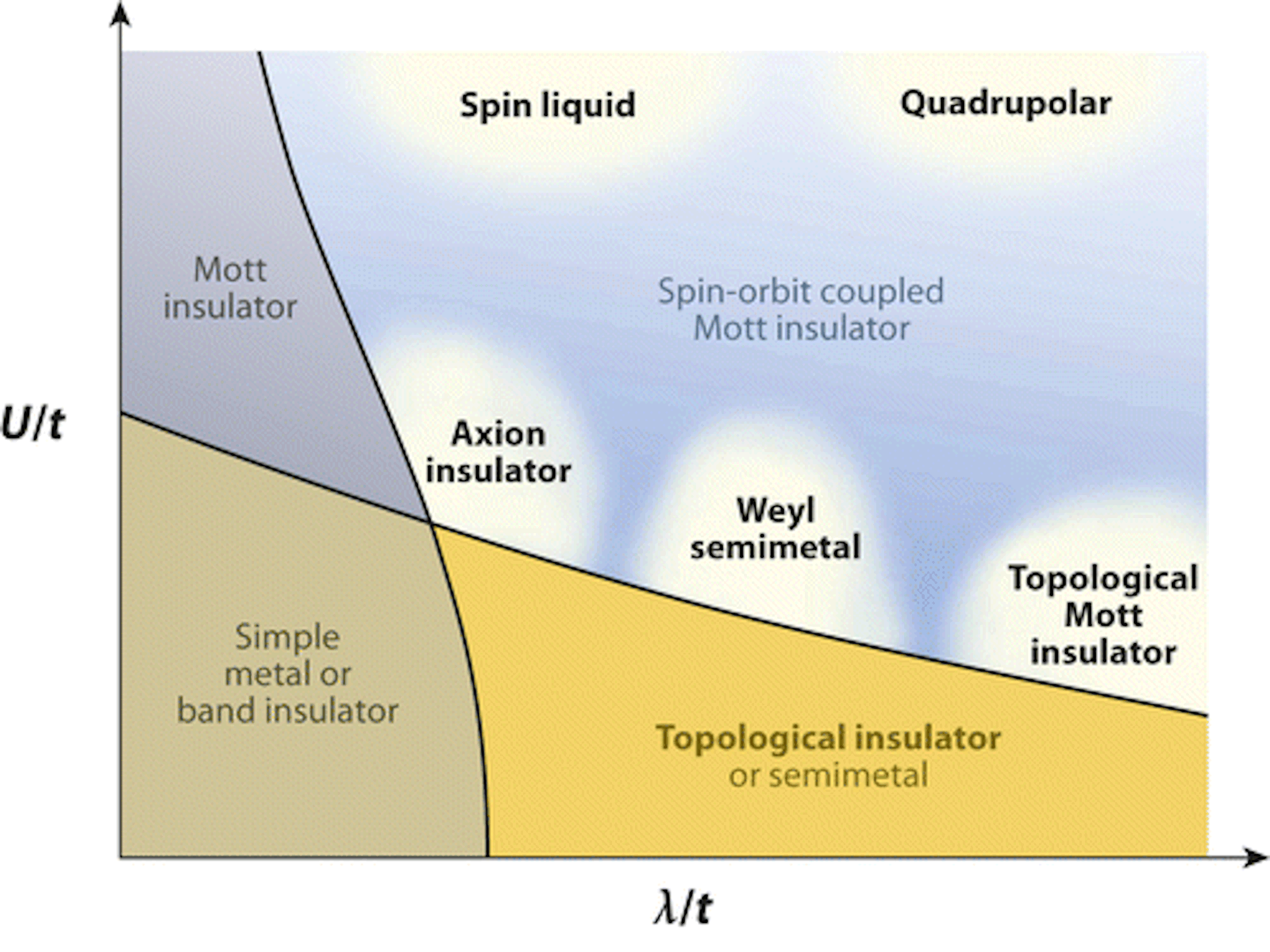}
    \caption{Sketch of a generic phase diagram for electronic materials in terms of the interaction strength \textit{U/t} and spin-orbit coupling \textit{$\lambda$/t}. Used with permission of \textit{Annual Reviews}, from Witczak-Krempa, et al. (2014) \cite{witczak-krempa_correlated_2014}; permission conveyed through Copyright Clearance Center, Inc.}
    \label{phasediagram}
\end{figure}

While we will focus primarily on challenges and opportunities in MBE synthesis of oxide quantum materials, it is worth defining the field in a broader sense. Much ink has been spilled on defining a ``quantum material." For a working definition of the field, we refer the reader to the United States Department of Energy \textit{Basic Research Needs Workshop on Quantum Materials for Energy Relevant Technology} from 2016 \cite{osti_1616509}. The workshop report emphasized that a variety of physical phenomena tied to quantum mechanics can be harnessed for future technologies. These include novel magnetic states, such as topological skyrmions, which may be useful for magnetic data storage. Novel superconducting materials that exhibit topological states are also of interest, as these are potential candidates for braided topological quantum computing systems \cite{das_sarma_proposal_2006}. An introduction to the field is also available through the review article of Cava, et al. \cite{cava_introduction_2021} and references therein. A common theme in many proposed systems is the need for strong spin-orbit coupling within the material to produce emergent topological states. A qualitative phase diagram for quantum matter that compares electron-electron repulsion \textit{U} and the atomic spin-orbit coupling \textit{$\lambda$} is shown in Figure \ref{phasediagram} \cite{witczak-krempa_correlated_2014}. Several oxide systems lie within these classes of materials, which we will highlight within this article. Strong spin-orbit coupling is present in atoms with high atomic number, such as 5d transition metals, making these elements particularly important for quantum materials. 

In this perspective article, we introduce readers to the MBE technique while also discussing trends and advances that will continue to improve our ability to synthesize ultra-high-quality quantum materials. We note the challenges associated with MBE growth in 4d and 5d transition metal systems and some of the developments that have enabled successful MBE synthesis of such materials. We have structured this article to offer discussion on some of the most exciting materials systems in the field today, including superconductors, such as nickelates and tantalates, and topological systems, such as iridates and ruthenates. Within these sections, we discuss how MBE has enabled some important breakthroughs and also how it has answered outstanding questions from previous research performed using other growth techniques. 

\subsection{Approaches to Synthesis}
For many years, complex oxides were primarily synthesized via pulsed laser deposition (PLD) from stoichiometric ceramic targets. This approach was first developed as a means to examine superconducting oxides, including bismuthates and cuprates\cite{christen_recent_2008}. It was initially thought to be a straightforward way to produce stoichiometric films, which is non-trivial for complex materials that do not exhibit an epitaxial growth window. Beyond superconducting materials, key breakthroughs came in the form of multiferroic BiFeO$_{3}$ thin films\cite{wang_epitaxial_2003}, interfacial 2D electron gases (2DEGs) in LaAlO$_{3}$/SrTiO$_{3}$ \cite{ohtomo_high-mobility_2004}, and numerous other systems. Over time, however, it became clear that material transfer was not uniformly stoichiometric and that growth conditions in PLD had significant impacts on resulting material properties \cite{breckenfeld_effect_2013,breckenfeld_effects_2014}. While PLD is still commonly employed for research in oxide films, including recent advances in superconducting nickelates \cite{li_superconductivity_2019}, more and more groups have begun employing molecular beam epitaxy as a way to grow films using thermalized adatoms that can be carefully controlled from elemental sources. 

Initial efforts in molecular beam epitaxy (MBE) of complex oxides focused primarily on 3d transition metals, beginning circa 1990. SrTiO$_{3}$ films were a primary focus for technique development in the emerging field since they can be grown homoepitaxially \cite{haeni_rheed_2000} and are also of interest for their prospects as high-k gate oxides in silicon devices \cite{mckee_crystalline_1998}. Additionally, efforts in cuprate film synthesis reflected the primary motivations of the era in condensed matter physics to better understand the high T$_{c}$ superconducting materials \cite{bozovic_atomic-level_1997}. Over time, numerous other materials of interest in the 3d block generated attention, including nickelates \cite{kumah_effect_2014, disa_orbital_2015}, manganites \cite{may_enhanced_2009, bhattacharya_metal-insulator_2008, monkman_quantum_2012}, ferrites \cite{ihlefeld_adsorption-controlled_2007, scafetta_band_2014, comes_interface_2016}, vanadates \cite{caspi_effect_2022}, cobaltates \cite{lee_strong_2019, andersen_layer-by-layer_2018}, and chromites \cite{chambers_band_2011, comes_probing_2017, lin_interface_2018, al-tawhid_two-dimensional_2021}. Additional efforts on band-insulating LaAlO$_{3}$ synthesis by MBE helped to resolve long-standing controversies in the field regarding the 2D electron gas at the LaAlO$_{3}$/SrTiO$_{3}$ interface \cite{qiao_epitaxial_2011, warusawithana_laalo3_2013, segal_x-ray_2009}. These works showed that stoichiometric films did not yield conductive interfaces or built-in electric fields, clarifying that the polar catastrophe model did not govern the behavior \cite{chambers_instability_2010}. As the field has come to better understand the 3d perovskite oxides, however, new challenges have emerged, with shifts in the field towards quantum materials that often involve refractory 4d or 5d transition metals on the B site. Because refractory elements generally cannot be grown from an effusion cell due to the extreme temperatures required for thermal evaporation, this has led to significant efforts to find alternative approaches for delivery of these cations.

\section{Developments in Oxide Molecular Beam Epitaxy}
In this section, we will discuss progress in the development of new ways to deliver transition metal cations. In general, in perovskite oxides (ABO$_3$), A site cations are straightforward to deliver from elemental source material via an effusion cell. However, B site ions provide much of the functionality for these materials. As we move into materials that exhibit quantum phenomena, we need the ability to repeatably and controllably deliver refractory elements that have historically been deposited via electron-beam evaporation within an MBE, as in the cases of SrRuO$_{3}$ \cite{nair_synthesis_2018} and SrIrO$_{3}$ \cite{nie_interplay_2015}. While electron-beam evaporation can in principle be viable with sufficiently precise atomic-absorption control \cite{du_self-corrected_2014}, we know of no MBE system that has implemented atomic-absorption in a reproducible fashion. Thus we focus our discussion on emerging techniques that offer an alternative to e-beam evaporation, as these hold greater promise for precise control of material quality.

One emerging approach to MBE synthesis that we will touch on only briefly is machine learning (ML) or artificial intelligence (AI) driven synthesis, which is still in its infancy. An enormous amount of data is available during the growth process, including in situ reflection high energy electron diffraction (RHEED), cell and sample stage temperatures, chamber pressures, source flux measurements, and other growth parameters. Post-growth characterization via X-ray diffraction, X-ray photoelectron spectroscopy, and electron and scanning probe microscopy can also provide key details on film quality. By recording full videos of RHEED patterns acquired during growth and combining these with log files of growth conditions and post-growth characterization, it is in principle possible to generate sufficient data to train neural networks that could ultimately provide feedback to enable AI-driven synthesis. To date, machine learning and data analytics efforts have focused primarily on either interpretation of RHEED videos to extract additional information \cite{provence_machine_2020,vasudevan_big-data_2014, suyolcu_engineering_2022} or ML-assisted optimization of growth conditions, such as in SrRuO$_{3}$ \cite{wakabayashi_machine-learning-assisted_2019}. These approaches take promising initial steps, but much work remains to be done in the area. We note that groups employing MBE in their work can contribute to the cause even if they do not develop ML algorithms themselves simply by recording full-length RHEED videos with corresponding metadata and log files for all of their growths so that others may use the videos in training of ML codes.

\begin{figure}
    \centering
    \includegraphics{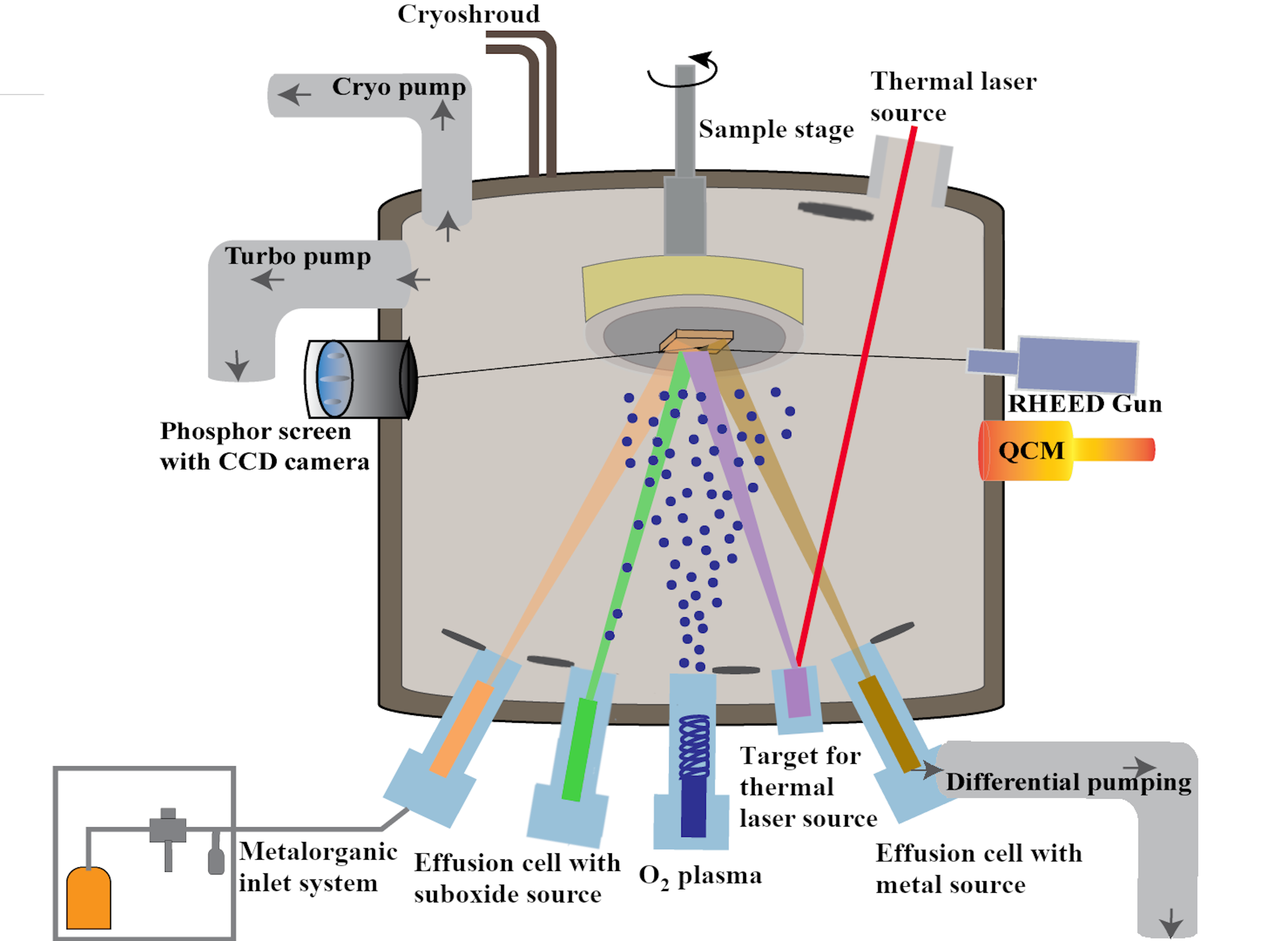}
    \caption{Schematic of current and future oxide MBE deposition techniques, including metalorganic, suboxide, thermal laser, and traditional metal source cell. Note that some metalorganic precursors can also be evaporated from an effusion cell. Differential pumping is often employed for metallic sources such as Sr to improve source stability or provide a means to quickly load new source material without venting the entire MBE chamber.}
    \label{combMBE}
\end{figure}

\begin{figure}
    \centering
    \includegraphics{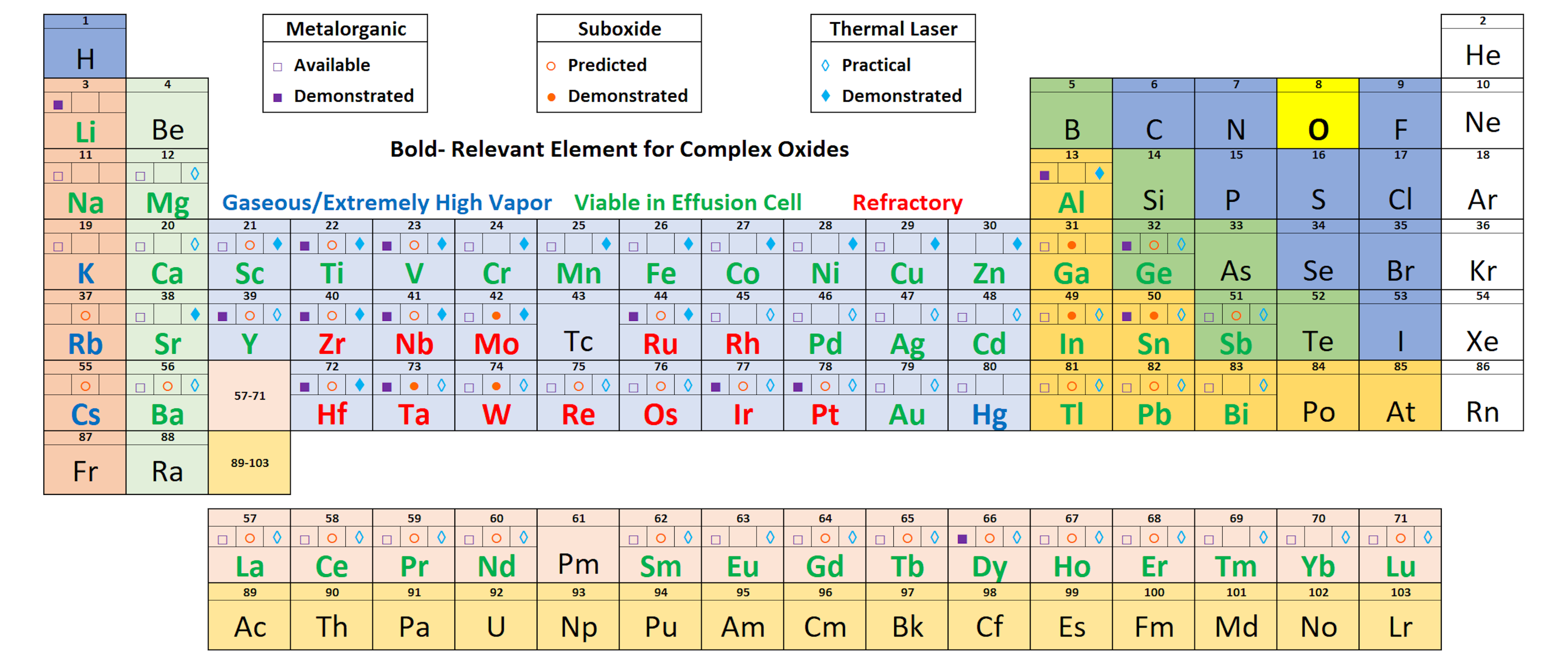 }
    \caption{Periodic table of elements relevant to oxide MBE. Growth techniques and future possibilities using emerging approaches to synthesis are demonstrated.}
    \label{periodictable}
\end{figure}

In the sections below, we discuss three new techniques for the delivery of refractory transition metals that are important for quantum materials. These include hybrid or metalorganic MBE, suboxide MBE, and thermal laser epitaxy. One can imagine a future where different sources are integrated into the same growth chamber to optimize the deposition for a particular class of materials. We show such a hypothetical system in Figure \ref{combMBE}. These techniques have been improving over the past decade, and many elements have already been incorporated using one or more approaches. A periodic table of the relevant elements for complex oxide films is shown in Figure \ref{periodictable}. This figure highlights the refractory elements, as well as high vapor pressure elements such as K, that have been historically difficult to implement in an MBE system via an effusion cell. It also shows the progress of the various techniques in depositing the elements of interest.

\subsection{Hybrid and Metalorganic MBE}
One avenue to delivery of metals that are challenging to evaporate in a conventional effusion cell is the use of a metalorganic precursor, many of which are now ubiquitous thanks to the widespread use of atomic layer deposition (ALD). While this approach is not new, dating back to the growth of metalorganic MBE (MOMBE) in the 1980s \cite{kuech_recent_1992}, it has only been commonly applied in oxide MBE since circa 2007 \cite{brahlek_frontiers_2018, jalan_growth_2009}. By redesigning an MBE chamber to accommodate a heated gas injector and developing a pressure-regulated feedback loop, metalorganic precursors can be delivered into an ultra-high vacuum environment. This approach has commonly been referred to as hybrid MBE, because the A site cations in a perovskite are still delivered from an effusion cell, while the B site cations are delivered through a metalorganic MBE source. For extensive details on the history of the approach, see the work of Brahlek et al \cite{brahlek_frontiers_2018}. Materials that have been synthesized include SrTiO$_{3}$ \cite{jalan_growth_2009, jalan_molecular_2009, son_epitaxial_2010}, rare-earth titanates \cite{xu_stoichiometry-driven_2014, xu_predictive_2016, moetakef_growth_2012}, SrVO$_{3}$ \cite{eaton_growth_2015, brahlek_accessing_2015}, BaSnO$_{3}$ \cite{prakash_adsorption-controlled_2017, prakash_wide_2017}, and SrNbO$_{3}$ \cite{thapa_surface_2022}, among others. 

Among the greatest accomplishments of hybrid MBE has been the demonstration of materials with some of the highest quality on record. SrTiO$_{3}$ films have been shown to exhibit an epitaxial growth window for cation stoichiometry \cite{jalan_molecular_2009}, analogous to what is seen in III-V materials grown by MBE. This has led to the highest reported carrier mobilities on record for SrTiO$_{3}$ films \cite{son_epitaxial_2010}. While it is still not entirely clear what surface chemistry produces the stoichiometric growth window, it is apparent that the titanium tetra-isopropoxide (TTIP) precursor preferentially adheres to a surface with a SrO termination and is more likely to desorb from a TiO$_{2}$-terminated surface \cite{brahlek_frontiers_2018,thapa_correlating_2021}. Stoichiometric growth windows have also been reported for SrVO$_{3}$ \cite{brahlek_accessing_2015} and BaSnO$_{3}$ films \cite{prakash_adsorption-controlled_2017}. 

Because stoichiometric control of complex oxide films has historically been such a challenge in conventional MBE, the availability of a growth window for these materials gives hybrid MBE a great deal of promise for quantum materials further down the periodic table. Perhaps the biggest challenge to date has been in the determination of suitable metalorganic precursors that span the constituent elements in the transition metal block. Difficulties related to the vapor pressure of the precursor, its melting temperature, and the radicals that are released after dissociation on the film surface all play a significant role in this selection. Specifically, ligands cannot contain elements that will contaminate a UHV chamber, such as Cl, F, or P, which are sometimes used in ALD processes. While most precursors contain some variation on hydrocarbon ligands that contain C, H, and O, in some cases a precursor that contains nitrogen-based ligands has proven to be more suitable, as in the case of SrNbO$_{3}$ growth using a (t-butylimido)tris(diethylamino) niobium (C$_{16}$H$_{39}$N$_{4}$Nb) precursor  \cite{thapa_surface_2022}. An analogous precursor exists for Ta, offering a route to the synthesis of tantalate films as well.

One alternative that overcomes the challenges of some liquid precursors that require pressure regulation has been to use a solid metalorganic source material. Groups have demonstrated this approach for Pt, Ru, and Ir using acetylacetonate (acac) ligands \cite{nunn_novel_2021, nunn_solid-source_2021, nair_engineering_2023}. These precursors are solid at room temperature and evaporate above 100 \degree C, making them viable in low temperature effusion cells. Several examples of this new technique will be discussed below in the sections on iridate and ruthenate films. 


\subsection{Suboxide MBE}
Recent advances in the growth of wide bandgap semiconductors such as Ga$_{2}$O$_{3}$ \cite{vogt_adsorption-controlled_2021} have led to spillover into the complex oxide film community and have generated new approaches for obtaining high quality films. New ideas have been applied to MBE growth of binary and ternary oxide semiconductors by using  metal-oxide sources rather than elemental metals. This approach relies on difference in vapor pressures between metallic sources and oxide sources. In many cases, a binary oxide of the form A$_{x}$O$_{y}$ will produce a higher flux of metal adatoms at the same effusion cell temperature than an elemental metallic source. This is particularly important for refractory elements that are difficult to evaporate from an effusion cell, such as 4d and 5d transition metals. Thermodynamic calculations by Adkison et al \cite{adkison_suitability_2020} showed that much of the periodic table is accessible through suboxide MBE, including low vapor-pressure transition metal elements such as Zr via ZrO, Nb via NbO$_{2}$, Mo via MoO$_{3}$, Ru via RuO$_{2}$, Hf via HfO, Ta via TaO$_{2}$, W via WO$_{3}$, Ir via IrO$_{2}$, and Pt via Pt$_{2}$O. To date, suboxide sources have been employed for the growth of MoO$_{3}$ films \cite{du_iso-oriented_2016} and WO$_{3}$ films \cite{li_crystallographic_2015} by MBE, and there have been limited reports of the use of suboxide MBE for transition metal perovskite oxides. We highlight the most prominent example, KTaO$_{3}$ \cite{schwaigert_molecular_2023}, below and note that oxides such as SrMoO$_{3}$ \cite{kuznetsova_growth_2023}, BaSnO$_{3}$ \cite{raghavan_high-mobility_2016} and Ba$_{n+1}$In$_{n}$O$_{2.5n+1}$ \cite{hensling_epitaxial_2022} have been successfully synthesized. This method has also been employed to catalyze surface reactions that results in higher growth rates \cite{Vogt2022}. As the oxide field moves towards a greater emphasis on quantum materials that employ 4d and 5d transition metals, we expect that suboxide MBE will fill a significant gap in the current synthesis capabilities.

\subsection{Thermal Laser MBE}
One additional approach to the delivery of refractory elements has emerged over the past 3 years that offers significant promise for the delivery of refractory elements in ultraclean conditions. Using a continuous-wave thermal infrared laser, refractory elements can be evaporated at high deposition rates, comparable to those attained by electron-beam evaporation \cite{smart_thermal_2021, braun_film_2019, kim_thermal_2021, kim_thermal_2022}. With this approach, the surface of a refractory element can be heated to the sublimation or evaporation temperature of the material while the rest of the material remains solid. Such a technique bridges the gap between electron-beam evaporation, which is difficult to control with sufficient precision for stoichiometric MBE growth, and effusion cells, which are not able to achieve sufficiently high temperatures. In particular, binary oxides of refractory 4d and 5d transition metals have been produced using this approach, including Nb, Hf, Zr, Ru, and Mo \cite{kim_thermal_2021}. Ongoing efforts in this area have focused on the development of careful temperature control such that laser power can be controlled via a feedback loop to maintain constant flux that is comparable to what can be achieved via an effusion cell. If appropriate control systems and growth chamber designs can be engineered, thermal laser MBE has a bright future for oxide and other quantum materials systems. 

In addition, the same thermal laser heating approach has been shown to produce pristine substrate surfaces through in situ annealing to ~1500 \degree C immediately before film growth \cite{braun_situ_2020}. This approach to sample heating in MBE can potentially remove the need for chemical pre-treatments for substrates and reduce outgassing from SiC heaters in oxide MBE systems. 

\subsection{Other Developments}

Additional developments in material processing techniques have provided a route to grow high-quality oxide films. Recently, it was demonstrated that kinetics of noble metals may be utilized to grow films in a fashion similar to adsorption control \cite{rimal_diffusion-assisted_2022}. In such a method, excess amount of a transition metal species can diffuse into the substrate through the film, while the film retains its stoichiometry. This proof-of-concept method was demonstrated in the growth of CuCrO$_{2}$, wherein excess Cu diffuses into an Al$_{2}$O$_{3}$ substrate due to its high diffusion kinetics. It was demonstrated that the CuCrO$_2$ film also acts as a filter that remains stoichiometric, yet helps excess Cu to diffuse through the substrate. Questions still remain on the nature of the diffused atoms, and whether a topotactic reaction occurs around the interface, but this method is likely to be useful in the case of materials involving other noble species where adsorption control is generally not feasible. Further studies on this behavior is likely to answer many details that leads to this behavior. 

A new method also combines laser in MBE. In this case, dubbed the hybrid PLD, pulsed laser sources, which are typically used in PLD, are used to to ablate binary oxide targets and generate reasonable fluxes of one species of atoms while traditional effusion cells are used for secondary species. This was recently demonstrated in the adsorption-controlled growth of KTaO$_3$ \cite{Kim2023_arxiv}. In this instance, a ceramic Ta$_2$O$_5$ target is ablated by a pulsed laser while a conventional effusion cell is used to generate excess K flux. It should be noted that the critical difference between this technique and thermal laser method is that in this method, flux is generated from a target is being \textit{ablated}, not heated as in thermal laser MBE. However, the ability to use ceramic targets may expand the possibility to use numerous types of low vapor pressure oxides as the molecular source.

Finally, another development in epitaxial film growth is the concept of remote epitaxy \cite{Kum2020,kim_remote_2022}. In this method, the film is grown on a substrate that has been convered by an atomically thin 2D material template. It was found that even though the substrate is covered by a layer of a 2D material, the film is commensurate with the substrate lattice. This method has been applied mainly to obtain free-standing oxide films, such that the film properties are not influenced by the substrate. However, one main challenge for oxide growth lies in the potential destruction of the 2D layer in activated oxygen or elevated temperatures. Hybrid MBE is therefore a promising avenue for growth of oxide films by remote epitaxy, because for growing materials such as SrTiO$_{3}$, the oxygen ligands in the precursor supply the necessary oxygen and the use of an additional oxygen source can be circumvented \cite{yoon_freestanding_2022}. The use of these membranes is now being investigated for various applications, including twisted bilayer structures \cite{Ying2022}.



\section{Magnetic and Ferroelectric Oxides}
MBE has been applied for growing many conventional magnetic and ferroelectric materials. The advantage of being able to span over five orders of magnitude of oxygen partial pressure, along with more powerful oxidants has allowed the growth of materials that are difficult to grow. Numerous ferromagnetic, antiferromagnetic and other magnetic materials have been investigated using MBE. The same can be said of ferroelectrics and multiferroics, and we refer to other review articles that have been published in this topic \cite{Chambers2000a,Mannhart2010,Ramesh2019,Bhattacharya2014}.

\section{Superconducting Oxides}

Superconductivity in oxides was first discovered in SrTiO$_{3}$ \cite{Schooley1964}, followed by Pb doped BaBiO$_{3}$ \cite{Sleight1975}. However, it was the discovery of high temperature superconductivity in cuprates \cite{Bednorz1986,Wu1987} that changed the direction of materials and condensed matter physics research. Subsequently, PVD methods such as PLD, sputtering, and MBE were modified to deposit and grow thin films of these wonder materials. MBE was identified as a powerful method to obtain high quality superconductors for both fundamental and applied research. Below, we discuss how MBE has been applied and adapted to some exemplary superconducting material systems. 

\subsection{Cuprates: The original high T$_c$ superconductor}
The potential and the development of oxide MBE can be traced to the growth of cuprate thin films \cite{Webb1987}. Precise control of doping and oxidation are some of the strong reasons that MBE is used to grow superconducting films and heterostructures. Extensive studies have been carried out on MBE-grown cuprates, such as La$_x$Sr$_{1-x}$CuO$_4$ (LSCO), Bi$_2$Sr$_2$CuO$_{6+x}$ (BSSCO), YBa$_2$Cu$_3$O$_{7-x}$ (YBCO) and variants \cite{baiutti_oxide_2018,yamamoto_epitaxial_2015}, often involving custom-designed MBE systems that allows fine-tuning of various subtle growth parameters \cite{Bozovic2001}. There are two main challenges in the growth on superconducting cuprates: oxidation and stoichiometry control. The issue with oxidation can be easily circumvented by using gas sources such as ozone or oxygen plasma. The more challenging issue arises from proper stoichiometry control. Superconductivity is due to the 2D copper oxide layers and since the properties of the films are highly dependent on doping, precise flux control is critical. Multiple in-situ methods such as a beam flux monitor, QCM, RHEED oscillations, and atomic absorption spectroscopy-based calibration provide high confidence in stoichiometry control of these materials. 

Recent progress on MBE-grown cuprates as been exemplified by heterostructures involving La$_2$CuO$_4$ \cite{suyolcu_octahedral_2017} and YBa$_2$Cu$_3$O$_{7-x}$ \cite{suyolcu_a-axis_2021}. In heterostructures of Sr$_2$CuO$_4$ and La$_2$CuO$_4$, STEM and EELS was used to obtain the spatial information on different CuO planes to help understand the localization of holes in superconducting heterostructures \cite{bonmassar_design_2023}. Further progress has also been made in the infinite-layer cuprates such as SrCuO$_2$. This phase of the cuprates are generally not as well studied as the others, since the thermodynamic stability is a big issue. Recently, detailed thermodynamic studies on the growth of the infinite layer cuprates was carried out, and plasma-assisted MBE was employed to successfully grow superconducting Ca$_{1-x}$Sr$_x$CuO$_2$ \cite{krockenberger_infinite-layer_2018}. Other groups have also been able to synthesize infinite layer cuprates using ozone \cite{Harter2015,Zhong2018}. 

The ease of doping and thickness control has resulted in ultra-high quality films and hetrerostructures allowing the investigation of the nature of the dimensionality of copper oxide \cite{Logvenov2009,Logvenov2013,Bozovic2016}. For more details on the synthesis-related details, we refer to recent review articles on the growth of cuprates using layer-by-layer MBE \cite{suyolcu_design_2020,xu_synthesis_2022}. 

As a final note, recent theories on twisted layers of cuprates have predicted novel states such as topological superconductivity \cite{haenel_incoherent_2022,tummuru_josephson_2022}, majorana modes \cite{margalit_chiral_2022} and pair density waves \cite{song_doping_2022}. Experiments are already ongoing on exfoliated layers of BSSCO \cite{zhao2021emergent} and the renewed interest on this old material system is certainly exciting from a synthesis perspective.

\subsection{Nickelates: Isoelectronic analogue to the cuprates}
Rare-earth nickelate thin films with the chemical formula RNiO$_{3}$ have been explored extensively over the past two decades due to the metallic nature of LaNiO$_{3}$ and the metal-insulator transition and antiferromagnetism present when other rare-earth elements are substituted, such as Nd, Sm, and Pr. For a review on the physics of perovskite nickelates with the wide range of phenomena they exhibit, readers are referred to the work of Catalano et al \cite{catalano_rare-earth_2018}. Much of this work was ultimately motivated by the analogy of the nickelates to cuprate superconductors, where the Ni$^{1+}$ ion is isoelectronic to the Cu$^{2+}$ ion. 

In general, perovskite nickelate thin films are a challenge to synthesize by MBE with ideal oxygen stoichiometry, due to the lower oxidation potential of late 3d transition metal ions such as Ni. Groups have found success with synthesis of LaNiO$_{3}$ using distilled ozone\cite{king_atomic-scale_2014} or oxygen plasma sources to enhance oxidation potential. Annealing after each unit cell of growth in the materials has also been shown to enhance the oxidation of the materials \cite{provence_machine_2020, may_onset_2009}. Various other \textit{in situ} synchrotron studies during growth have focused on structural recovery of epitaxial films during the initial growth stages on mixed-termination substrates \cite{li_self-healing_2022, yan_situ_2020}. However, MBE-grown films have provided excellent platforms for studies of nickelate physics, including the effects of surface termination \cite{kumah_effect_2014} and thickness \cite{king_atomic-scale_2014} on the metallicity of LaNiO$_{3}$. These results suggest that surface effects, due to the combined contributions of octahedral distortions and surface chemistry, play a significant role in transport in nickelate films. In the case of the highest quality LaNiO$_{3}$ films grown via MBE, evidence of antiferromagnetism has been seen below 1 K, with a strong dependence on film disorder in the form of cation off-stoichiometry ($\sim$2\%) and differing oxidation conditions \cite{liu_observation_2020}. Given the similarities between nickelate and cuprate physics, significant efforts were undertaken to use interfaces to engineer orbital polarization in LaNiO$_{3}$ to drive the material closer to the cuprate electronic configuration \cite{disa_research_2015, disa_orbital_2015}. Collectively, our understanding of nickelate physics has been significantly advanced through careful studies of MBE-grown films. 

A significant breakthrough within the nickelate research field came with the first observation of superconductivity in nickelate films grown by PLD \cite{li_superconductivity_2019}. In this work, Li et al reported that a layered phase of Nd$_{0.8}$Sr$_{0.2}$NiO$_{2}$ could be produced via chemical reduction of Nd$_{0.8}$Sr$_{0.2}$NiO$_{3}$ using a topotactic phase transition induced by annealing in CaH$_{2}$ powder in a tube furnace. This layered phase is analogous to the SrCuO$_{2}$ phase in the cuprates, but is unstable in atmospheric conditions and required a SrTiO$_{3}$ capping layer in the initial work \cite{li_superconductivity_2019}. Further work has been performed on PLD-grown infinite-layer nickelates with this structure, as is discussed in recent reviews by Nomura and Arita\cite{nomura_superconductivity_2022} and Bernardini et al \cite{bernardini_thin-film_2022}. Notably, a recent work in PLD nickelate films argued that hydrogen incorporation into the films during the CaH$_{2}$ annealing step was responsible for the superconductivity \cite{ding_critical_2023}. This result raises important questions as to the effects of the synthesis process, particularly the annealing step. Efforts to explore alternative synthesis avenues to achieve the same effective Ni charge state as the infinite-layer structure are critical to understand the mechanisms of the superconductivity in nickelate films.


Two particularly noteworthy approaches have been pursued through MBE film growth in superconducting nickelates. In the first case, Pan et al. synthesized layered Ruddelsden-Popper (RP) nickelates that are analogous to the layered cuprate superconductors \cite{pan_superconductivity_2022}. Nd$_{6}$Ni$_{5}$O$_{16}$ films were grown by MBE and then reduced to Nd$_{6}$Ni$_{5}$O$_{12}$ using the same CaH$_{2}$ reduction process. These films have an average Ni formal charge of Ni$^{1.2+}$, producing a 3$d^{8.8}$ electronic configuration that matches the configuration for the layered cuprates. Such layered materials would be a significant challenge to grow by PLD, but by alternating between NdO and NiO$_{2}$ layers via shuttering the group was able to stabilize the RP phase. After reduction, the films showed maximum T$_{c}$ values of $\sim$13 K, with a broad superconducting transition that does not reach zero resistance until $\sim$1 K. These results are notable for the lack of a SrTiO$_{3}$ capping layer due to the greater atmospheric stability of the RP phase in comparison to the infinite-layer phase. Also notable is the broad phase transition in comparison to the infinite-layer structures. Given the complexity of the RP phase and the propensity of such films to form stacking faults that reduce the uniformity of the layered structure \cite{ferenc_segedin_limits_2023,pan_synthesis_2022}, it suggests that structural inhomogeneities play a significant role in the superconducting behavior of the materials. Additionally, hydrogen may also be present within the films thanks to the CaH$_{2}$ reduction process. 

\begin{figure}
    \centering
    \includegraphics{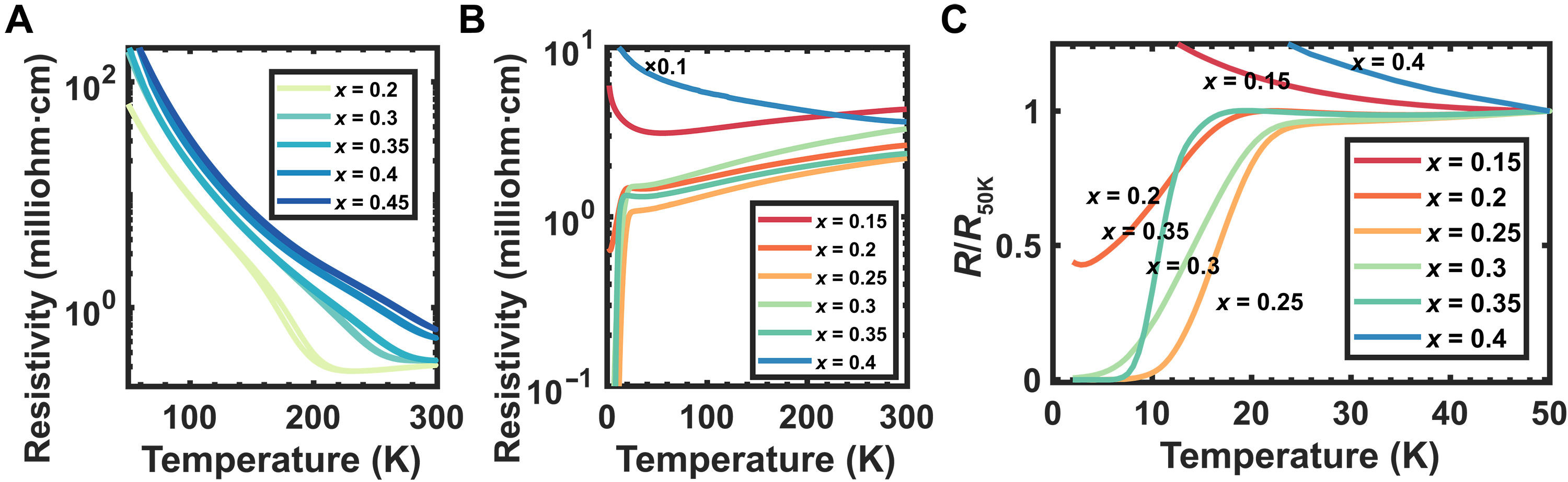}
    \caption{Superconductivity in Nd$_{1-x}$Eu$_{x}$NiO$_{2}$ films grown with reducing Al capping layer. (A) Resistivity-versus-temperature behavior of Nd$_{1-x}$Eu$_{x}$NiO$_{3}$ samples shows a metal-to-insulator transition at a temperature that depends on Eu concentration x. (B) Resistivity-versus-temperature curves (2 to 300 K) of Nd$_{1-x}$Eu$_{x}$NiO$_{2}$ samples show superconductivity in the doping level $0.2 {<} x  {<} 0.35$. The resistivity for the $x = 0.4$ sample is multiplied by a factor of 0.1. Resistivity-versus-temperature curves between 2 and 50 K are plotted in (C) and are normalized at 50 K. Figure and caption are reproduced in whole from Wei et al. \cite{wei_superconducting_2023} under the \href{https://creativecommons.org/licenses/by/4.0/}{Creative Commons Attribution License 4.0 (CC BY)}.}
    \label{nickelates}
\end{figure}

Recent studies have particularly focused on reduction of nickelate films without CaH$_{2}$ using metallic aluminum capping layers \cite{wei_superconducting_2023, wei_solid_2023}. In these works, the authors grew epitaxial nickelate films followed by Al capping layers without air exposure. The Al film is highly reducing to the underlying nickelate material, driving the valence from the stoichiometric perovskite or Ruddelsden-Popper phase to the reduced phases. In the case of Ruddelsden-Popper films with the initial stoichiometry of Nd$_{6}$Ni$_{5}$O$_{16}$, the reduced films did not exhibit superconductivity in the Nd$_{6}$Ni$_{5}$O$_{12}$ phase \cite{wei_solid_2023}, contrary to the reports for MBE-grown films synthesized via CaH$_{2}$ reduction \cite{pan_superconductivity_2022}. Interestingly, infinite-layer Nd$_{1-x}$Eu$_{x}$NiO$_{2}$ films synthesized in this manner were superconducting and exhibit  T$_{c}$ values of 21 K for $x = 0.25$ \cite{wei_superconducting_2023}, higher than any previous reports for MBE or PLD-grown films. Transport data for these films is shown in Figure \ref{nickelates}. Clearly, there are still numerous questions left to address in regards to the origins of superconductivity in nickelate materials and the role of hydrogen, but MBE film growth has proven to be quite valuable thus far.

\subsection{Tantalates: A 5d superconductor}
Superconductivity was first discovered in KTaO$_3$ surfaces by field-effect induced doping \cite{ueno_discovery_2011}, and work over the past few years examining tantalate films and heterostructures has also shown the analogy of KTaO$_{3}$ to SrTiO$_{3}$ in terms of its tunability at epitaxial interfaces. These studies were first inspired by the work of Liu et al. on EuO/KTaO$_{3}$ heterostructures grown by MBE on KTaO$_{3}$ (111) substrates, where superconductivity with T$_{c}$ of $\sim$1.5 K was reported \cite{Liu2021}. This superconducting phase transition was attributed to electron doping of the KTaO$_{3}$ crystal due to creation of oxygen vacancies by the overgrowth of EuO by MBE and LaAlO$_{3}$ by PLD. Subsequent work by the same group showed that the critical temperature could be tuned electrostatically by varying the carrier concentration in the interfacial 2DEG and explained how phonon-mediated pairing drove the superconductivity at the (111)-oriented interfaces but not in (100)-type interfaces \cite{liu_tunable_2023}. Other works have employed magnanite capping layers  grown by MBE on KTaO$_{3}$ (111) substrates to achieve similar superconducting behavior \cite{arnault_anisotropic_2023,al-tawhid_enhanced_2023}. These films showed that the higher spin-orbit coupling in KTaO$_{3}$ in comparison to the analogous superconducting titanate interfaces led to more complex dependence on the in-plane direction of current flow and direction of applied magnetic field \cite{arnault_anisotropic_2023}. In general, the promise of superconducting materials that exhibit strong spin-orbit coupling such as KTaO$_{3}$ leaves a great deal of room for future exploration. 

\begin{figure}
\centering
\includegraphics[scale=0.7]{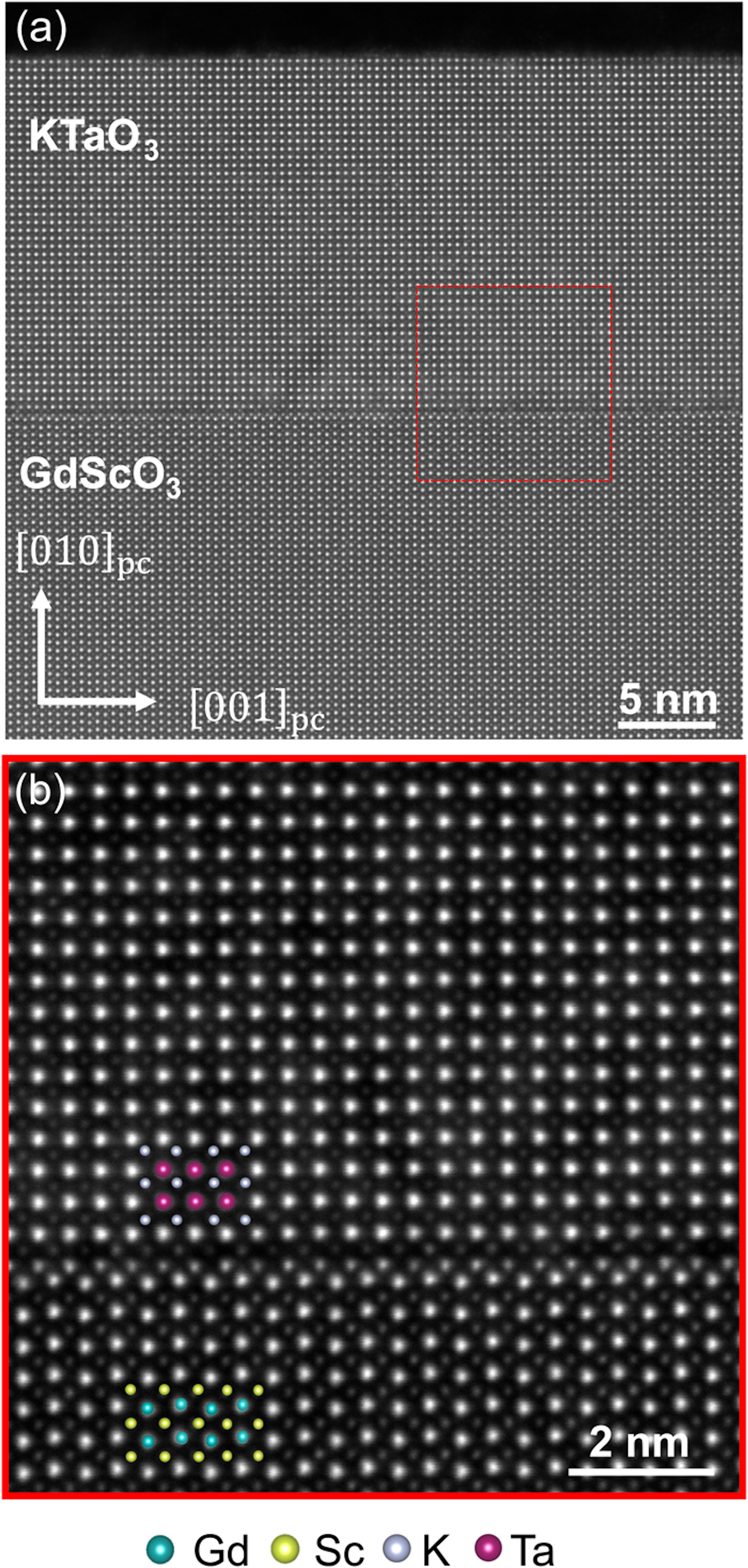}
    \caption{(a) Cross-sectional HAADF-STEM images of the same 18 nm thick KTaO$_{3}$ film grown on a GdScO$_{3}$ (110) substrate with an effusion cell containing Ta$_{2}$O$_{5}$. (b) Higher magnification HAADF-STEM image of the KTaO$_{3}$/GdScO$_{3}$ interface showing the bilayer of the intermixed metal ions. Figure and caption reprinted with permission from Schwaigert, et al \cite{schwaigert_molecular_2023}. Copyright 2023, American Vacuum Society.}
    \label{KTaO3}
\end{figure}

One of the biggest challenges in the field is the need to develop ways to synthesize tantalate films by MBE. In the works from the preceding paragraph, KTaO$_{3}$ single crystal substrates were used for film growth and only non-refractory materials were deposited by MBE. While others have grown KTaO$_{3}$ by PLD \cite{gupta_ktao3new_2022}, only recently have attempts to synthesize the material by MBE been reported \cite{schwaigert_molecular_2023}. In that work, K, which has an extremely high vapor pressure at near-ambient conditions in elemental form, was delivered from an effusion cell that contained an In-K alloy. Meanwhile, Ta was delivered in the form of TaO$_{2}$ by suboxide MBE from a Ta$_{2}$O$_{5}$ source and as Ta metal from an e-beam source. A STEM image from a film grown from the suboxide source is shown in Figure \ref{KTaO3}. This is an important demonstration of suboxide MBE and is the first step towards more complex tantalate heterostructures that can be grown by MBE.

\subsection{Non transition metal oxides}
The perovskite BaPb$_{1-x}$Bi$_x$O$_3$ was the second oxide that was shown to be a superconductor, exhibiting a relatively high T$_C$ of about 13 K \cite{Sleight1975}. Further research led to the discovery of superconductivity in Ba$_{1-x}$K$_x$SbO$_3$ (with a lower T$_C$) \cite{batlogg_superconductivity_1989} and Ba$_{1-x}$K$_x$BiO$_3$ with a higher T$_C$ of over 30 K \cite{mattheiss_superconductivity_1988,cava_superconductivity_1988}. 

Studies of MBE-grown thin films of these materials  soon followed \cite{hellman_molecular_1989} and more efforts were made to investigate potential superconductivity in heterostructures involving the bismuthates \cite{bozovic_quest_2020}. Recently, superconductors based on group III and IV metals have gained a new interest, driven by the possibility of topological phenomena due to their high spin-orbit coupling. First principles calculations has also predicted a topological insulator state in doped BaBiO$_3$ surface \cite{yan_large-energy-gap_2013}, showing the possibility for complex phases to exist in this system. Research on this system involving thin films and heterostructures is still ongoing \cite{harris_superconductivity-localization_2018,harris_charge_2020} and new insights on the interplay of strong spin-orbit coupling, superconductivity and symmetry breaking have been found. It should be pointed out that one challenge in the growth and heterostructuring of the bismuthates also lies in the charge disproportionation of Bi. Since Bi cannot stabilize in Bi$^{+4}$ state, it forms Ba$_2$Bi$^{+3}$Bi$^{+5}$O$_6$ structure involving distortions, octahedral tilts, and a charge density wave \cite{sleight_bismuthates_2015}. 

Although superconductivity was discovered in Sb-doped BaPbO$_{3}$, the first report on only Sb based material was reported only recently, with T$_C$ reaching 15 K \cite{kim_superconductivity_2022}. A charge density wave was also found, similar to the bismuthates. These old, yet relatively unexplored systems provide incentives and new avenues for novel properties and phases in a non transition metal oxide.


\section{Topological Phenomena in Oxides}
Topological phenomena are ubiquitous in modern day condensed matter and materials physics. Topological effects have been invoked since the discovery of quantum Hall effects in high-mobility 2DEG semiconductors \cite{avron_topological_2003}. The prediction and discovery of the quantum spin-hall effect \cite{Murakami2004,Konig2012} and topological insulators accelerated the research on topological materials. In general, most studies on topological materials are focused on high spin-orbit coupled materials with broken symmetry, which results in electronic bands crossing or inversion \cite{Bansil2016,Narang2020}. Competition between different energies provides new routes to novel phases \cite{Witczak-Krempa2014}. 

Although oxides do not garner much attention during discussion of topological materials, a search of oxides in topological databases \cite{Vergniory2019,topologicalQuantumChemistry} shows that over 25\% of candidates are predicted to be either semimetals or topological insulators. In fact, detailed studies have predicted topological band crossings in numerous oxides, and many oxides are predicted to host Dirac and Weyl fermions, including SrRuO$_{3}$ \cite{chen_weyl_2013,gu_overview_2022}, EuTiO$_{3}$ \cite{takahashi_control_2009,takahashi_anomalous_2018}, RbCr$_{4}$O$_{8}$ \cite{xia_robust_2019}, Ag$_{2}$BiO$_{3}$ \cite{he_tunable_2018}, pyrochlore iridates \cite{bzdusek_weyl_2015} and others \cite{uchida_topological_2018}. SrRuO$_{3}$ has been widely discussed in the literature as a Weyl material, with some experiments showing signatures for such a case \cite{itoh_weyl_2016,takiguchi_quantum_2020,toyoda_weyl_2022}. 

Thin film heterostructures are also predicted to host topological Dirac and Weyl nodes, for example in rutile heterostructures \cite{pardo_half-metallic_2009,lado_quantum_2016} and perovskites\cite{xiao_interface_2011}. Epitaxial films of SrRuO$_{3}$ \cite{gu_overview_2022}  and La-doped EuTiO$_{3}$ \cite{takahashi_anomalous_2018} have been studied, and signatures of topological behavior has been observed. Similarly, real space topological textures called skyrmions are also being actively investigated in oxide films and heterostructures. We refer the readers to a recent review which provides examples of these oxide materials \cite{uchida_topological_2018}. Below we discuss two of the typical oxide materials systems in terms of topological behavior: ruthenates and iridates.

\subsection{Ruthenates: Topological textures and superconductivity}
Real space topology, namely topological spin textures (skyrmions and its variants) have been widely investigated, especially in the spintronics community \cite{He2022}. For recent review on these topological spin textures, see Junquera et al \cite{Junquera2023} and Tokura et al \cite{tokura_magnetic_2021}. Generally, inversion symmetry needs to be broken for skyrmions to exist, and the antisymmetric Dzyaloshinskii-Moriya interaction is invoked in explaining many features of skyrmion-like behavior in SrRuO$_{3}$ films and heterostructures. Various numerical and experimental methods are being used to predict or measure the effects of this interaction \cite{kuepferling_measuring_2023}. 

Transport based measurements, namely involving the topological Hall effect, have claimed to prove the existence of skyrmions in various oxides \cite{Ohuchi2015,Yun2018,ohuchi_electric-field_2018,Sohn_stable_2021}. SrRuO$_{3}$ is an exemplary oxide that is frequently cited in literature for hosting skyrmions, with many experiments linking the topological Hall effect to the existence of skyrmions. However, recent experiments using engineered films and heterostructures, many using MBE, show that factors such as sample inhomogeneity and competing anomalous Hall effect and Berry phases are responsible for these hump-like features \cite{Kimbell2020,Kim2020,Skoropata2021,Wang2020,schreiber_model_2023,manjeshwar_adsorption-controlled_2023}. Therefore, careful assessment of material properties is required and these features in anomalous Hall effect may not be necessarily related to topological Hall effect caused by skyrmions or similar textures. 

MBE-grown SrRuO$_3$ films have been used for investigating multiple physical phenomena. High residual resistivity ratio (RRR), which is tied to the higher quality and lower defects, has been achieved for films grown using e-beam evaporated MBE \cite{nair_synthesis_2018} and hybrid metalorganic MBE \cite{manjeshwar_adsorption-controlled_2023}. As a ferromagnetic conductor with high spin-orbit coupling, SrRuO$_3$ has been investigated from spintronic and topological perspectives, and a recent review by Gu et al provides an good overview \cite{gu_overview_2022}.

\begin{figure}
\centering
    \includegraphics[scale=0.9]{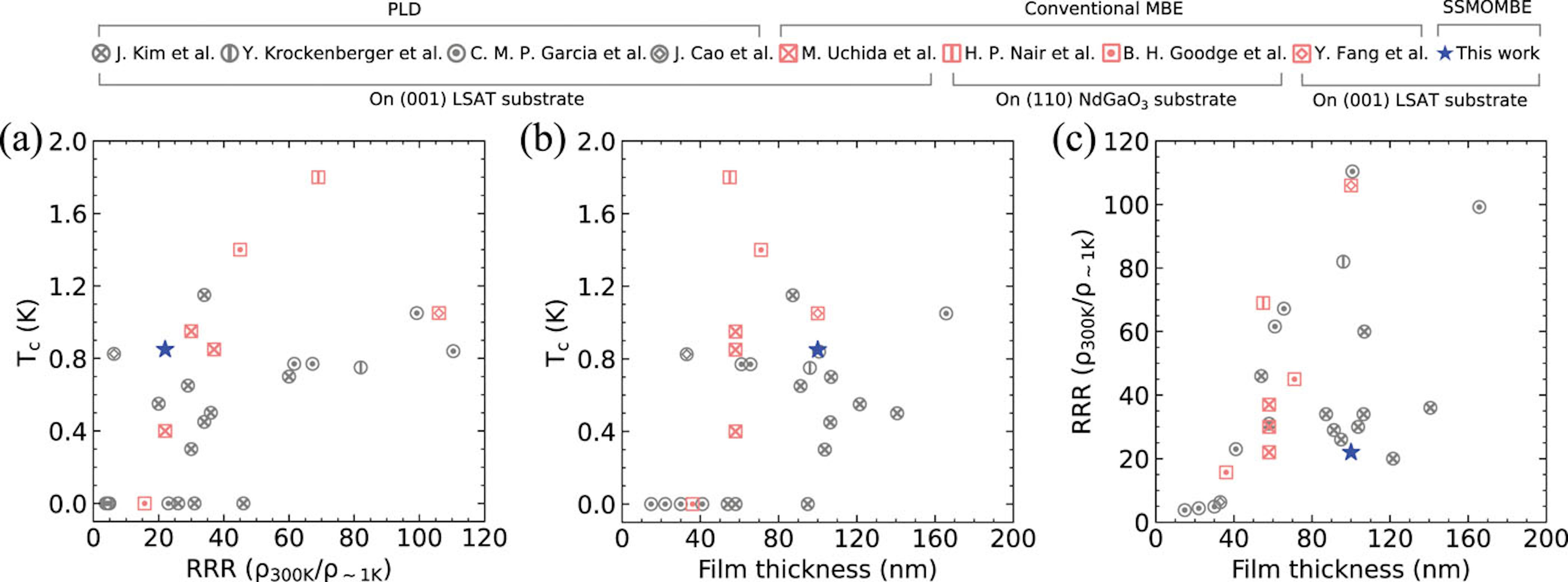}
    \caption{Superconducting transition temperature ($T_c$) of Sr$_{2}$RuO$_{4}$ thin films as a function of (a) residual resistivity ratio (RRR) and (b) film thickness. (c) RRR vs film thickness of Sr$_{2}$RuO$_{4}$. The row at the top of the legends indicates the technique used to grow the films, and the row at the bottom indicates the type of substrate used. The $T_c$ plotted here is the midpoint of superconductivity onset temperature and zero-resistance temperature. Figure and excerpt of caption taken from Choudary et al (solid-source metalorganic MBE (SSMOMBE), star symbol) \cite{choudhary_growing_2023} under the \href{https://creativecommons.org/licenses/by/4.0/}{Creative Commons Attribution License 4.0 (CC BY)} with other data extracted from references therein \cite{krockenberger_growth_2010,  kim_superconducting_2021, nair_demystifying_2018, uchida_molecular_2017, cao_enhanced_2016,goodge_disentangling_2022,garcia_pair_2020,fang_quantum_2021}.}
    \label{Sr2RuO4}
\end{figure}

Because of the unusual p-wave spin-triplet superconductivity believed to be present in the Ruddelsden-Popper Sr$_{2}$RuO$_{4}$ phase\cite{ishida_spin-triplet_1998, mackenzie_even_2017}, these materials have also been an area of significant interest for epitaxial films for many years as possible candidates for topological quantum computation \cite{das_sarma_proposal_2006}. Because of the sensitivity of spin-triplet superconductors to impurities and defects, efforts to achieve superconducting films via PLD took many years with few reports of success and T$_{c}$ values up to 1.0 K \cite{krockenberger_growth_2010, kim_superconducting_2021}. Subsequent work using MBE via electron-beam evaporation \cite{nair_demystifying_2018, uchida_molecular_2017} of Ru later produced films with T$_{c}$ values up to 1.8 K on NdGaO$_{3}$ substrates \cite{nair_demystifying_2018}. This work clarified the importance of both film purity and choice of substrate to yield superconducting films. Recent work in metal-organic MBE using a Ru(acac)$_{3}$ solid precursor demonstrated that superconductivity could also be achieved via this novel method, with a reported T$_{c}$ of 0.8 K \cite{choudhary_growing_2023}. A comparison of the critical temperatures for various film growth techniques, RRR values, and substrates is shown in Figure \ref{Sr2RuO4}. Further work with thermal laser MBE to synthesize superconducting films from a metallic Ru source is also ongoing \cite{faeth2023NAMBE}.

\subsection{Iridates: Dirac and Weyl materials}
Pyrochlore and honeycomb lattice based phases of iridates, along with heterostructures, have been predicted to host novel electronic states \cite{Witczak-Krempa2014,bzdusek_weyl_2015}. Most of these materials have been studied in bulk single-crystal form and thin films are slowly being developed. Other phases such as Sr$_{2}$IrO$_{4}$ has also been predicted to be a superconductor, although no experimental verification has been realized. Similarly, SrIrO$_{3}$-based (hetero)structures have been predicted to be topological insulators \cite{carter_semimetal_2012}.

Thin films have been fabricated using various methods, and we would like to refer to some recent reviews and articles that focus on iridate thin films \cite{kim_perspective_2022,liu_magnetic_2021}. Heterostructures based on iridates and other transition metal oxides are being actively investigated that allows for fine tuning of interfacial effects.  Particularly, SrIrO$_{3}$ is interfaced with magnetic complex oxides such as LaMnO$_{3}$ and SrRuO$_{3}$ that may result in interfacial topological textures \cite{liu_interfacial_2019,skoropata_interfacial_2020,liu_controllable_2022,jeong_correlated_2023}. At the interface of this high spin-orbit coupled material, charge transfer, combined with symmetry breaking and hybridization effects, may allow the development of novel phases. Sharp interfaces require the growth of defect-free, high-quality films and new methods are actively being investigated to grow these materials. 

\begin{figure}
\centering
    \includegraphics[scale=0.9]{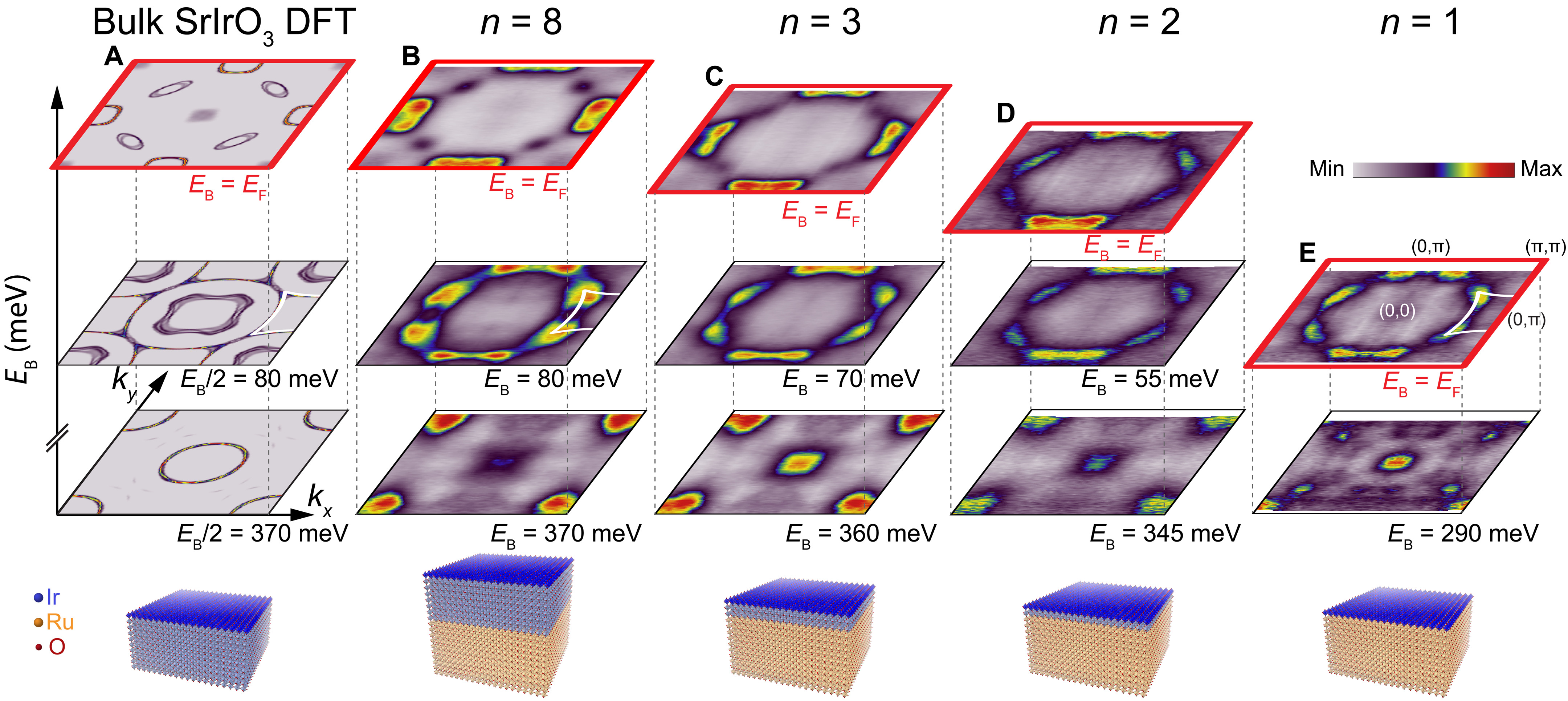}
    \caption{Constant energy intensity ARPES maps of (SrIrO$_{3}$)$_{n}$/(SrRuO$_{3}$)$_{20}$ heterostructures. (A) Density functional theory (DFT) calculations of constant energy intensity maps for bulk SrIrO$_{3}$ at a variety of binding energies, $k_z = 0.5 \pi/c$. The binding energies are renormalized by a factor of 2 to account for many body effects described in the text. The conversion between experiment and DFT is $E_B = E_{DFT}/2$ - 0.035 eV. (B to E) ARPES intensity maps for the \textit{n} = 8, 3, 2, and 1 unit cell heterostructures, with the top row showing the Fermi surface at \textit{E$_{B}$} = \textit{E$_{F}$} highlighted in red. The square hole pockets are highlighted in white as a guide to the eye in the 80 meV maps of the DFT, \textit{n} = 8 unit cell and the Fermi surface of the \textit{n} = 1 unit cell, demonstrating that the map consists purely of hole pockets. Figure and caption adapted from Nelson, et al. \cite{nelson_interfacial_2022} under the \href{https://creativecommons.org/licenses/by/4.0/}{Creative Commons Attribution License 4.0 (CC BY)}.}
    \label{SIO-SRO}
\end{figure}

The extremely low vapor pressure of Ir is the main bottleneck for MBE growth of the iridates, and e-beam evaporation of Ir was typically used \cite{nie_interplay_2015,kawasaki_evolution_2016}. The challenge in using e-beam evaporator is flux stability and source degradation. To overcome those issues, recent developments in hybrid MBE \cite{choudhary_semi-metallic_2022} has provided new routes. So far, MBE has been used for the growth of the binary oxide IrO$_2$ \cite{kawasaki_evolution_2016} and the Ruddlesden-Popper phases SrIrO$_3$ and Sr$_2$IrO$_4$ \cite{nie_interplay_2015,kawasaki_evolution_2016,choudhary_semi-metallic_2022,nelson_mott_2020}. Heterostructures involving SrIrO$_3$ and SrRuO$_3$ were also investigated, and the role of charge transfer was investigated using ARPES \cite{nelson_interfacial_2022}. Their work showed that Ir donates a fraction of an electron to Ru at the interface, changing the Fermi surface of the materials. ARPES measurements of these samples are shown in Figure \ref{SIO-SRO}.

A challenge in MBE synthesis of iridates lies with the pyrochlores. Other methods such as PLD has been used to grow many pyrochlore iridates and signatures of topological phases have been found in these films \cite{liu_magnetic_2021}. It is thus appealing to investigate how pyrochlores grow, and determine the potential of heterostructures involving these materials.

\section{Future Directions and Conclusion}
There is a large role of oxide MBE in driving fundamental and applied research, and the prospects look promising. The development of new MBE methods that were outlined above has allowed the exploration of multitude of phases. Much work remains to be done in refining these techniques and implementing them into research-scale MBE systems that are widely available within academic and government labs and industrial research and development. 

It must be noted that one of the big challenges in MBE growth of some of the outlined materials is the availability of proper substrates. There are already numerous substrates that allow lattice and symmetry matching for many transition metal oxides. Currently, substrates with (pseudo)cubic lattice constant of $3.74-4.00$ {\AA} are available commercially. Buffering with different materials can also be used for lattice matching, but may damage the functional properties of a given film or heterostructure. As new materials with novel properties are added to the collection, new types of substrates are needed. Collaboration between thin film and bulk crystal growers enables development of new substrate materials that can potentially be used for materials of the future. For example, synthesis of (La,Nd)(Lu,Sc)O$_{3}$ bulk crystal provides greater lattice constant for growing the wide-bandgap material BaSnO$_{3}$ which has a much higher lattice constant than the available substrates \cite{Guguschev2020}. Development of substrates for pyrochlore phases such as the iridates described above is also underway \cite{Bovo_2017, C7CE00942A}.

Another challenge is inherent to materials with hexagonal symmetry. Most of the commercially available oxide substrates have cubic crystal structure, and although cutting along the hexagonal (111) axis is possible, this cut in a perovskite has high surface energy resulting in uneven surfaces. Chemical treatments sometimes allow flattening of these surfaces in some substrates, but there is no single method that works for all. Since numerous topological materials have hexagonal symmetry, development of new hexagonal substrates is a necessity. As an example, CuFeO$_2$ single crystals have been developed as substrate candidates for delafossite films \cite{WOLFF2020125426}. 

Additionally, there is a long-standing challenge in directly probing the electronic structure of  topological materials. While indirect evidence may be obtained by electronic transport \cite{Hu2019}, first principles calculations are typically employed to predict if a material is topological \cite{Bradlyn2017}, and experimental evidence can be provided via techniques such as ARPES which can directly probe the band structure \cite{Lu2012}. Recently, MBE has been integral in determining the topological properties of materials. By in-situ transfer from growth to analysis chambers with XPS, ARPES, and/or STM capabilities \cite{thapa_probing_2021}, band structure and interfacial phenomena can be directly probed without exposing the surface to atmosphere. Similar challenges are encountered in materials with real-space topology, especially in cases where skyrmions are predicted to be confined at interfaces. New developments in integrated experimental probes with real-space sensitivity are thus required for properly determining the topological properties of thin films and heterostructures. These may include spin-resolved ARPES in a photoemission electron microscope (PEEM) or spin-resolved STM, either of which can be integrated with an MBE.

To summarize, we have discussed how advances in oxide thin film growth by MBE have pushed the field in new directions for quantum materials synthesis. MBE has been used to answer outstanding questions and design new layered materials that would not be possible via other techniques. Developments in new approaches to cation delivery, including the use of metalorganic precursors, suboxide sources, and thermal laser heating, have enabled better control of growth of refractory 4d and 5d elements with high spin-orbit coupling that are widely used in quantum materials. These new techniques will push the field forward in coming years as we continue to explore exciting new functional behavior of complex oxides for quantum applications. 

\section{Acknowledgements}
Work at Auburn University was supported by the U.S. Department of Energy (DOE), Office of Science, Basic Energy Sciences (BES), under Award DE-SC0023478 (GR, topological materials) and by the National Science Foundation (NSF) under Award DMR-2045993 (RBC, superconducting materials).

\section{References}

\bibliographystyle{iopart-num}  
\bibliography{references_GR}


\pagebreak

\end{document}